# Excitation of surface plexciton wave at interface of a metal and a columnar thin film infiltrated with J-aggregate dyes

F. Babaei [*], M. Rostami

*Department of Physics, University of Qom, Qom, Iran*

*)Email: fbabaei@qom.ac.ir

## ABSTRACT

In this paper, we report theoretical investigation on excitation of surface plexciton wave at interface of a metal and a columnar thin film infiltrated with J-aggregate dyes using transfer matrix method in Kretschmann configuration. The results reveal the regime of the plasmon - exciton interaction can change from weak to strong by tuning structural parameters. We find Rabi splitting energies between 204 -378 meV corresponding to the time period 11-20 fs which includes to the fast energy transfer from surface plasmon polaritons to excitons. The phase speed and propagation length of surface plexcitonic waves were in the range of 0.4 to 0.9c and 0.4 to 6 μm. The time-averaged Poynting vector of surface plexciton waves shows the localization of them at interface of plasmonic and excitonic mediums.

## I. INTRODUCTION

The interaction of plasmons and excitons is studied in new field of condensed matter physics in the name of plexcitonics [1]. A plexcitonic system is a combination of a plasmonic medium(metal) and an excitonic medium(semiconductor) [2]. When plasmon frequency of metal ($\omega_{SPP}$) is close to exciton frequency of semiconductor($\omega_X$)[3,4], coupling between plasmon and exciton creates a new quasiparticle that it is called plexciton. Plexcitons are new optical hybrid states including low polariton (LP) and upper polariton(UP) branches [5]. Rabi splitting $\Omega_R$ is minimum energy between LP and UP polariton branches at zero detuning frequency δ=0 ($\delta = \omega_{SPP} - \omega_X$)[6].The plasmon – exciton interaction enhances optical Stark effect due to amplifying optical nonlinearities[7].The enhanced optical Stark effect can induce spin splitting of electrons, even can manipulate spin in hybrid nanostructures[8] and even may be used in spintronic in the future. The hybrid nanostructures consisting Rabi oscillations can apply for ultrafast optical switching in active plasmonic devices [9]. Rabi splitting phenomena in plexcitonic systems are potential candidates for nanolasers, biosensors, and single-photon switches [10].

The porosity of columnar thin films grown by oblique angle deposition [11,12] allows us to fill them with excitonic molecules. In spite of various works which have been done on plasmon-exciton interaction, to the best of our knowledge, this is the first work that studies the propagation of surface plexciton wave at interface of a metal and a columnar thin film infiltrated with J-aggregate dyes. Here, metal and porous columnar thin film were considered Silver(Ag) and Magnesium fluoride(MgF$_2$), respectively. We assumed excitons (Frankel excitons at room temperature) of a cyanine dye [5,5′,6,6′-tetrachloro-di-(4-sulfobutyl) benzimidazolocarbocyanine(TDBC) have been self-assembled as J-aggregate due to its strong dipole moment [13]. We theoretically investigate the plasmon-exciton coupling at planar interface of a metal and a columnar thin film infiltrated with J-aggregate dyes using transfer matrix method in Kretschmann configuration.

## II. COMPUTATIONAL DETAILS

Fig.1 shows an illustration our geometry for excitation of surface plexcitonic wave in the Kretschmann configuration. Suppose that the region $0 \leq z \leq d_1$ is occupied by an isotropic homogenous metal as plasmonic medium with relative permittivity $\varepsilon_{met}$. The region $d_1 \leq z \leq t$ is considered as excitonic medium, where $t = d_2 \sin\chi$, while the regions $z \leq 0$ and $z \geq t$ are respectively occupied by a prism with relative permittivity $\varepsilon_1 = n_1^2$ and an air medium with refractive index n$_2$ =1. Let us assume that this set exposed by p- polarized plane wave from the bottom of the prism at an angle $\theta$ to the z- axis in xz-plane. The phasors of incident, reflected and transmitted electric fields are given:

$$\begin{aligned}
\underline{E}_{inc}(\underline{r}) &= [\underline{P}_{inc}]e^{i[k\ x+k_0\ n_1\ z\ \cos\theta\ ]}, & z \leq 0 \\
\underline{E}_{ref}(\underline{r}) &= [\underline{S}\ r_S + \underline{P}_{ref}\ r_P]e^{i[k\ x-k_0\ n_1\ z\ \cos\theta\ ]}, & z \leq 0 \\
\underline{E}_{tr}(\underline{r}) &= [\underline{S}\ t_S + \underline{P}_{tr}\ t_P]e^{i[k\ x+k_0\ n_2(z-d_1)\ \cos\theta'\ ]}, & z \geq t
\end{aligned} \qquad (1)$$

The phasor of the magnetic field in any region is given as $\underline{H}(\underline{r}) = (i\omega\mu_0)^{-1}\nabla \times \underline{E}(\underline{r})$, where $(r_S, r_P)$ and $(t_S, t_P)$ are the amplitudes of reflected and transmitted waves with s- and p-linear polarizations and also $k = k_0 n_1 \sin\theta$. It must be mentioned here, a time dependence $\exp(i\omega t)$ is implicit, with $\omega$ denoting the angular frequency. The free-space wave number, the free-space wavelength, and the intrinsic impedance of free space are denoted by $k_0 = \omega\sqrt{\varepsilon_0\mu_0}$, $\lambda_0 = 2\pi/k_0$ and $\eta_0 = \sqrt{\mu_0/\varepsilon_0}$, respectively, with $\mu_0$ and $\varepsilon_0$ being the permeability and permittivity of free space, respectively. Vectors are underlined once, dyadics are underlined twice, the asterisk denotes the complex conjugate and Cartesian unit vectors are identified as $\underline{u}_{x,y,z}$.

The unit vectors for linear polarization normal and parallel to the incident plane, $\underline{S}$ and $\underline{P}_{inc,ref,tr}$, respectively are defined as:

$$\begin{aligned}
\underline{S} &= \underline{u}_y \\
\underline{P}_{inc} &= -\underline{u}_x \cos\theta + \underline{u}_z \sin\theta \\
\underline{P}_{ref} &= \underline{u}_x \cos\theta + \underline{u}_z \sin\theta \\
\underline{P}_{tr} &= -\underline{u}_x \cos\theta' + \underline{u}_z \sin\theta'
\end{aligned} \qquad (2)$$

, where $\sin\theta' = (n_1/n_2)\sin\theta$, $\cos\theta' = +\sqrt{1-\sin^2\theta'}$.

The reflectance and transmittance amplitudes can be obtained, using the continuity of the tangential components of electrical and magnetic fields at interfaces and solving the algebraic matrix equation:

$$\begin{bmatrix} t_S \\ t_P \\ 0 \\ 0 \end{bmatrix} = [\underline{\underline{K}}(\theta')]^{-1}[\underline{\underline{M}}_{plas}][\underline{\underline{M}}_{exc}][\underline{\underline{K}}(\theta)]\begin{bmatrix} 0 \\ 1 \\ r_S \\ r_P \end{bmatrix} \qquad (3),$$

The transfer matrix of the excitonic $[\underline{\underline{M}}_{exc}]$ and plasmonic $[\underline{\underline{M}}_{plas}]$ mediums can be obtained by solving the source-free Maxwell curl postulates in that regions:

$$\nabla \times \underline{E}(\underline{r}) = i\omega \underline{B}(\underline{r})$$
$$\nabla \times \underline{H}(\underline{r}) = -i\omega \underline{D}(\underline{r})$$
(4),

where D and B are the displacement electric and the magnetic induction fields, respectively. The constitutive relations for D and B are as:

$$\underline{D}(\underline{r}) = \varepsilon_0 \underline{\underline{\varepsilon}}_{exc} . \underline{E}(\underline{r})$$
$$\underline{B}(\underline{r}) = \mu_0 \underline{H}(\underline{r})$$
(5)

in excitonic medium and

$$\underline{D}(\underline{r}) = \varepsilon_0 \varepsilon_{met} \underline{E}(\underline{r})$$
$$\underline{B}(\underline{r}) = \mu_0 \underline{H}(\underline{r})$$
(6)

in plasmonic mediums. The nonhomogeneous dielectric permittivity $\underline{\underline{\varepsilon}}_{exc}$ for excitonic medium is defined as:

$$\underline{\underline{\varepsilon}}_{exc} = \underline{\underline{S}}_y(\chi) . \underline{\underline{\varepsilon}}^e . \underline{\underline{S}}_y^T(\chi)$$
(7),

where the superscript T indicates to the transpose of a dyadic and χ is tilt angle of the nanocolumns of the excitonic medium.

The effective relative permittivity and tilt dyadic are respectively as:

$$\underline{\underline{\varepsilon}}^e = \varepsilon_a^e \underline{u}_z \underline{u}_z + \varepsilon_b^e \underline{u}_x \underline{u}_x + \varepsilon_c^e \underline{u}_y \underline{u}_y$$
$$\underline{\underline{S}}_y = (\underline{u}_x \underline{u}_x + \underline{u}_z \underline{u}_z)\cos\chi + (\underline{u}_z \underline{u}_x - \underline{u}_x \underline{u}_z)\sin\chi + \underline{u}_y \underline{u}_y$$
(8),

where $\varepsilon_{a,b,c}^e$ are the effective relative permittivity scalars of the excitonic medium.

The dielectric permittivities of porous columnar thin film (MgF$_2$ + air) $\varepsilon_{a,b,c}(\lambda, f_v)$ were obtained using Bruggeman homogenization method [14]. Fig.S1 in the supplementary material shows the refractive indexes of Magnesium fluoride, Silver, J-aggregate TDBC dyes and the dielectric permittivity scalars $\varepsilon_{a,b,c}(\lambda, f_v)$ of porous MgF$_2$ columnar thin film at different void volume fractions($f_v$). The effective dielectric permittivities scalars of the porous columnar thin film infiltrated with J-aggregate dyes with volume fraction $f_j$ ($f_j$<1) were obtained by [15]

$$\varepsilon^e_{a,b,c}(\lambda, f_v, f_j) = \varepsilon_{a,b,c}(\lambda, f_v)\,[1 + \frac{f_j}{\varepsilon_{a,b,c}(\lambda, f_v)/[\varepsilon_j(\lambda) - \varepsilon_{a,b,c}(\lambda, f_v)] + (1-f_j)/3}] \qquad (9).$$

With considering electromagnetic fields in excitonic and plasmonic regions are as follows:

$$\underline{E}(r) = \underline{e}(z)\,e^{ikx}$$
$$\underline{H}(r) = \underline{h}(z)\,e^{ikx} \qquad (10),$$

and using the Eqs.4-6, one can obtain four ordinary differential equations and two algebraic equations. The two algebraic equations for $e_z$ and $h_z$ in excitonic and plasmonic mediums are respectively as follows:

$$e_z(z) = \frac{\dfrac{(\varepsilon_a - \varepsilon_b)}{2} e_x(z) \sin 2\chi - \dfrac{k}{\varepsilon_0 \omega} h_y(z)}{B}$$

$$h_z(z) = \frac{k}{\mu_0 \omega} e_y(z) \qquad (11),$$

where $B = (\varepsilon_a \cos^2 \chi + \varepsilon_b \sin^2 \chi)$.

$$e_z(z) = -\frac{k}{\varepsilon_0 \omega \varepsilon_{met}} h_y(z)$$

$$h_z(z) = \frac{k}{\mu_0 \omega} e_y(z) \qquad (12).$$

The four ordinary differential equations can be sorted as a matrix equation:

$$\frac{d}{dz}[\underline{f}(z)] = i[\underline{\underline{P}}(z)][\underline{f}(z)] \qquad (13),$$

where $[\underline{f}(z)] = [e_x(z)\ e_y(z)\ h_x(z)\ h_y(z)]^T$ is a column vector. The $[\underline{\underline{P}}_{exc}(z)]$ is a $4\times 4$ matrix and the elements of it for excitonic region are :

$$P_{11} = \frac{k(\varepsilon_a - \varepsilon_b)\sin 2\chi}{2B},\ P_{14} = \mu_0\omega - \frac{k^2}{\varepsilon_0 \omega B},\ P_{23} = -\mu_0\omega,$$

$$P_{32} = \frac{k^2}{\mu_0 \omega} - \varepsilon_0 \omega \varepsilon_c,\ P_{41} = \varepsilon_0\omega\frac{\varepsilon_a \varepsilon_b}{B},\ P_{44} = P_{11} \qquad (14),$$

and the other elements are zero. The elements of $[\underline{\underline{P}}_{plas}(z)]$ in plasmonic region are :

$$P_{14} = \mu_0\omega - \frac{k^2}{\varepsilon_0\omega\varepsilon_{met}}, \; P_{23} = -\mu_0\omega, \; P_{32} = \frac{k^2}{\mu_0\omega} - \varepsilon_0\omega\varepsilon_{met}, \; P_{41} = \varepsilon_0\omega\varepsilon_{met} \quad (15),$$

and the else elements are zero.

Now, easily the transfer matrix of excitonic and plasmonic mediums can be introduced as $[\underline{\underline{M}}]_{exc} = e^{i[\underline{\underline{P}}_{exc}]t}$ and $[\underline{\underline{M}}]_{plas} = e^{i[\underline{\underline{P}}_{plas}]d_1}$, respectively.

By rewriting Eqs.1 as $[\underline{f}(z=0_-)] = [\underline{\underline{K}}(\theta)][0 \; 1 \; r_S \; r_P]^T$ in incident medium(including incident and reflected electric fields ) and $[\underline{f}(z=t_+)] = [\underline{\underline{K}}(\theta_{tr})][t_S \; t_P \; 0 \; 0]^T$ in transmitted medium, the nonzero elements of $[\underline{\underline{K}}(\theta)]$ and $[\underline{\underline{K}}(\theta')]$ respectively are:

$$K_{12} = -K_{14} = -\cos\theta, \; K_{21} = K_{23} = 1, \; K_{31} = -K_{33} = \frac{n_1}{\eta_0}K_{12}, \; K_{42} = K_{44} = \frac{-n_1}{\eta_0} \quad (16),$$

and

$$K_{12} = -K_{14} = -\cos\theta', \; K_{21} = K_{23} = 1, \; K_{31} = -K_{33} = \frac{n_2}{\eta_0}K_{12}, \; K_{42} = K_{44} = \frac{-n_2}{\eta_0} \quad (17).$$

Using the obtained amplitudes in Eq. 3, then we can calculate the reflectance and transmittance coefficients as $r_{iP} = r_i$, $t_{iP} = t_i$, where $i = s, p$. The optical absorption for p- linear polarization is calculated as $A_P = 1 - \sum_{i=s,p}(R_{iP} + T_{iP})$, where $R_{iP} = |r_{iP}|^2$ (reflection) and $T_{iP} = (n_2 \text{Re}(\cos\theta')/(n_1 \cos\theta))|t_{iP}|^2$ (transmission) so that Re( ) is the real part of the quantity given in the parenthesis. Here, we considered only p- linear polarization in all simulations because there are not significant peaks for excitation of surface plexcitonic wave in s- linear polarization.

### III. RESULTS AND DISCUSSIONS

Figure 2 depicts absorption density ($A_P$) as functions of wavelength and incident angle for excitation of surface plasmon polariton(SPP), surface exciton(X) and surface plexciton(PLX). The considered parameters in Figs.2a1-2a3 and Figs.2b1-2b3 were respectively $d_1$= 40 nm, $d_2$ = 800 nm, $n_1$=2.57, $n_2$ =1, $\chi$ =25 °, $f_j$ = 0.052, $f_v$ = 0.748 and $d_1$= 40 nm, $d_2$ = 800 nm, $n_1$=1.52, $n_2$ =1, $\chi$ =25 °, $f_j$ = 0.1, $f_v$ =

0.1. Vertical lines refer to where $\omega_{SPP} = \omega_X$. The angular position of zero detuning frequency($\delta=0$) in Fig.2a3 and Fig.2b3 occurs at $\theta_\delta$ =25. 47° and $\theta_\delta$ =65. 45°, respectively, where they are shown with horizontal lines in plots. In experimental works, reaching to the zero detuning frequency is difficult due to the limitations of measuring instruments. Herein, we scanned wavelength and incident angle with increment 1nm and 0.01°, respectively. It is well known that there is a characteristic anticrossing between LP and UP branches of polaritons at $\delta=0$ in plexciton plots. Comparison between plexciton plots showed that the splitting energy between LP and UP branches at zero detuning frequency in Fig.2b3 is less than Fig.2a3. It must be mentioned here, in depiction of density plots for bare SPP and bare X, the volume fraction of J-aggregate dyes ($f_j$) and thickness of metal($d_1$) were zero and zero, respectively (Figs.2a1,2b1 and Figs.2a2,2b2). In fact, when frequency of the plasmon sates equals to frequency of the exciton states, the energy states of plexcitonic system splits(see Fig.2) to $\omega_+ = \omega_X + g$ and $\omega_- = \omega_X - g$ (g is coupling energy) on the basis of the coupled oscillator model(COM) framework [16].

Optical absorption as a function of photon energy at different volume fractions of air and J-aggregate dyes with fixed parameters $d_1$= 40 nm, $d_2$ = 800 nm, $n_1$=2.57, $n_2$ =1, $\chi$ =25 ° at zero detuning frequency $\delta=0$ for excitation of surface X, SPP and PLX waves at interface of plasmon and exciton mediums are given in Fig.3. The zero detuning frequency occurred in Figs.3a -3d at incident angles 25.47,27.54,29.91,32.46 degrees, respectively. Thus, the angular position of zero detuning frequency shifts to higher incident angles, because the optical index of exciton medium increases with the increasing fraction of host medium $f_s$=1-($f_j$+$f_v$) (MgF$_2$). It is clear that the splitting energy between LP and UP decreases by increasing of fraction of J-aggregate dyes. This result contrasts with what was previously reported by Balci et al. [17]. Although, the optical density of bare excitons in host medium increases with increasing of the concentration of TDBC molecules (see Fig.3). However, there is a difference between our theoretical work and Balci et al. experimental work can be related to the structural difference between the idealized theoretical model and that obtained in experimental work. In our work, the exciton medium is a composite with three components: dielectric columnar thin film(MgF$_2$), air and J-aggregate dyes. Here, the competition between fractions of host medium and sum fractions of air and J-aggregates dyes is decisive in the separation energy of LP and UP branches. In viewing to results from Fig.3d to Fig.3a despite the fact that the J-aggregate fraction is reduced but the sum fractions of air and J-aggregates dyes are increased. Then, the accumulation of air shells on exciton molecules strengths the dipole moment at interface and therefore the polariton branches are separated from each other (see Figs.3a-3d).

We followed the same calculations for variations of the thickness of metal $d_1$ (Fig.S2), the length of exciton medium $d_2$(Fig.S3), the tilt angle of nanocolumns of exciton medium $\chi$ (Fig.S4) and the refractive index of prism $n_1$(Fig.S5). Also, the plots of zero detuning angle $\theta_\delta$ for different structural parameters are given in Fig.4. The linewidth of SPP peak decreases with the plasmonic layer thickness in Fig. S2. At first the separation between the polaritonic branches increased for thickness of metal from 30 to 40nm and then decreased from 50 to 60nm. The angular position of zero detuning frequency was almost constant at 32°. If the thickness of metal is less than the saturated thickness, a high percentage of incident light is transmitted and plasmons do not have enough time for coupling to excitons. Also, if the thickness of metal is more than the saturated thickness, the energy of photons is dissipated in the metal and the efficiency of the SPP excitation decreases at interface due to the increasing of tunneling distance. Then, at a saturated of plasmonic layer arises a large separation between LP and UP branches and it depends on considered structure. It is clear from both Figs.S3 and S4 the splitting energy of polaritonic branches increased with the length and tilt angle of nanocolumns of exciton medium. Because the thickness of exciton medium is directly proportional to $d_2$ and $\chi$ as $d_2\sin\chi$ (see Fig.1). In both figures the angular position of zero detuning frequency was about 32°. We found that the angular position of zero detuning frequency shifted to higher incident angles with decreasing the refractive index of prism (see Fig.S5). Because the SPP peaks for coupling to excitons at $\omega_{SPP} = \omega_X$ moved to higher incident angles. In Fig.S5, the separation energy between polaritonic branches was almost unchanged.

For more details based on analysis the COM framework, the Rabi splitting energy $\Omega_R = 2g = \sqrt{(\omega_+ - \omega_-)^2 + 0.25(\gamma_{SPP} - \gamma_X)^2}$ [6,10] and splitting energy $\Delta\omega = \omega_+ - \omega_-$ are shown in Fig.4,where $\gamma_{SPP}$ and $\gamma_X$ were the full width half maximum (FWHM) of plasmons and excitons, respectively. It is seen that the Rabi splitting energy and splitting energy both with increasing fraction of J-aggregate decreased in Fig.4a. The difference between $\Omega_R$ and $\Delta\omega$ is the result of the FWHMs of plasmons and excitons. Except for $f_j=0.1$, where there is an intermediate regime $\gamma_{SPP} > 2g > \gamma_X$, for the rest of the values of $f_j$ there are strong coupling regimes between plasmons and excitons $2g > (\gamma_{SPP}, \gamma_X)$. We found that at first $\Omega_R$ and $\Delta\omega$ with plasmonic layer thickness respectively were decreased and increased in Fig.4b. Because the FWHM of plasmons becomes less from 30 to 40 nm (see Fig.S2). After 40 nm, the Rabi splitting energy and splitting energy were equal due to $\gamma_{SPP} = \gamma_X$. In our work, there is a change regime from intermediate to strong by the thickness of metal. In Figs.4c and 4d, both $\Omega_R$ and $\Delta\omega$ increased with the length and tilt angle of nanocolumns of excitonic medium. The obtained results revealed that there exists an intermediate regime except that the regime for $\chi=35,45°$ was strong. We found that the coupling between plasmons and excitons located in intermediate regime in Fig.4e. Also, the Rabi splitting energy decreased but the splitting energy was almost fixed (the difference was only 2-5 meV). In this work, the time period $T_R = 2\pi/\Omega_R$ for Rabi oscillation was in range 11-20 fs so that this time indicates to the fast energy transfer between plasmons and excitons. In addition, we think that different types of regimes for coupling plasmons and excitons are available even may reached to weak coupling regime $2g < (\gamma_{SPP}, \gamma_X)$ by adjusting structural parameters. We did not follow this regime in simulations.

In order to obtain wavenumber of surface plexcitonic wave, we depicted optical absorption as a function of incident angle at fixed wavelengths (see Figs.S6-S10 in supplementary material). These wavelengths $\lambda_{LP}$ and $\lambda_{UP}$ were the same peaks of PLX plot as LP and UP in Fig.3 and Figs.S2-S5. From $Ap(\theta)$ in Figs.S6-S10 is clear that the peaks for LP and UP branches occurred near zero detuning angle $\theta_\delta$. Deviation from $\theta_\delta$ obtained in Fig.4 indicates to broadening peaks. There are the other peaks in $Ap(\theta)$ in Figs.S6-S10 that they can be related to waveguide modes or cavity resonances of structure. We considered wavenumber of surface plexcitonic wave for LP($ksp_{LP}$) and UP($ksp_{UP}$) branches as $ksp_{LP,UP} = k_{LP,UP} n_1 \sin\theta + i k_{LP,UP} n_1 \sin\gamma_{LP,UP}$, where θ is angular position of peaks in Ap(θ), $k_{LP,Up} = 2\pi/\lambda_{LP,UP}$, $\gamma_{LP}$ and $\gamma_{UP}$ were the FWHMs in Ap(θ) for LP and UP branches. This method is used in Kretschmann configuration for propagation of surface waves at interface of two mediums [18,19]. The relative phase speed and the propagation length of surface plexcitonic wave introduced as $\frac{v}{c}\bigg|_{LP,UP} = \frac{k_{LP,UP}}{\text{Re}[ksp_{LP,UP}]}$ and $L_{LP,UP} = \frac{1}{\text{Im}[ksp_{LP,UP}]}$, respectively. These quantities for both branches of surface plextonic waves are shown in Fig.5. We found in Fig.5a that the relative phase speed and the propagation length for both LP and UP branches of surface plexcitonic wave decreased with the fraction of J-aggregate dyes. The maximum value for v/c = 0.9 and L=6 μm occurred for LP branch at $f_j$=0. 052. It is seen in Fig.5b that at first v/c increased then was constant for LP branch, while it was fixed for UP branch with the plasmonic layer thickness. The propagation length at interface increased for both branches with variation of thickness plasmonic layer. In Figs.5c,5d, the v/c for both branches at first decreased then was constant, while L increased with the length and tilt angle of nanocolumns of excitonic medium. Also Fig.5e revealed that the relative phase speed and the propagation length for both branches increased with the refractive index of prism. Generally, any change in the relative phase speed and the propagation length of the surface plexcitonic wave is due to changes of angular position, wavelength and FWHM of peaks. In all plots of Fig.5, the relative phase speed and the propagation length for LP branch is higher than UP branch. On the other hand, the LP plexcitonic surface wave is long lived while the other branch is short lived.

The localization of surface plexcitonic wave at interface of plasmonic and excitonic mediums was shown only for two samples in Fig.6 by depiction of the Cartesian components of time-averaged Poynting vector $|p_{x,z}(z)|$ along the z axis when a surface LP- and UP- plexciton wave was excited at interface. The used structural parameters were same to Fig.2. We considered the time-averaged Poynting vector as $\underline{P} = \frac{1}{2}\text{Re}[\underline{E} \times \underline{H}^*]$ in calculations. The penetration depth(pd) was obtained as 1/e of the intensity of $P_z$ in plasmonic and excitonic mediums that those values for LP and UP branches were $pd_{LP,plas} = 11 nm$, $pd_{LP,exc} = 181 nm$ (Fig.6a1), $pd_{UP,plas} = 14 nm$, $pd_{UP,exc} = 127 nm$ (Fig.6a2), $pd_{LP,plas} = 13 nm$, $pd_{LP,exc} = 136 nm$ (Fig.6b1) and $pd_{UP,plas} = 13 nm$, $pd_{UP,exc} = 90 nm$ (Fig.6b2). In all plots of Fig.6 were observed that the surface plexcitonic waves dissipated in plasmonic and excitonic mediums in the normal to interface for both LP and UP polaritonic branches. Therefore, the energy of photons transferred to quasiparticle plexcitons so that they were localized at plasmonic/excitonic interface.

## IV.   CONCLUSION

In conclusion, a detailed theoretical study of the plasmon-exciton interaction is presented. The dependence of the Rabi splitting energy, zero detuning angle, the relative phase speed and the propagation length of the surface plexcitonic wave to structural parameters are reported. We found a range for Rabi splitting energy between 204-378 meV from intermediate to strong coupling regime. The strong coupling regime causes to fast energy transfer between plasmons and excitons. The results show that the Rabi splitting energy and zero detuning angle with the sum factions of air and J-aggregate dyes in porous columnar thin film increased and decreased, respectively. The zero detuning angle was almost fixed at 32° by plasmonic layer thickness, the thickness and tilt angle of nanocolumns of the exciton medium. The angular position of zero detuning frequency moved to higher incident angles with decreasing the refractive index of prism due to shift the SPP peaks to higher incident angles for coupling to excitons at $\omega_{SPP} = \omega_X$. At onset of plasmonic layer the $\Omega_R$ decreases due to broadening plasmons at detuning frequency then the Rabi splitting energy equals to the splitting energy. It is obtained that the Rabi splitting energy increased with the length and tilt angle of nanocolumns of excitonic medium and also the Rabi splitting energy decreased with the refractive index of prism. The results of the relative phase speed and the propagation length for surface plexcitonic waves reveal that the access to v = 0.9 c and L=6μm are possible for LP polariton branch. In our work, the LP plexcitonic surface wave was long lived while the UP branch was short lived. Finally, the results of the time-averaged Poynting vector show localization of surface plexcitonic waves for both polaritonic branches at interface of plasmonic and excitonic mediums.

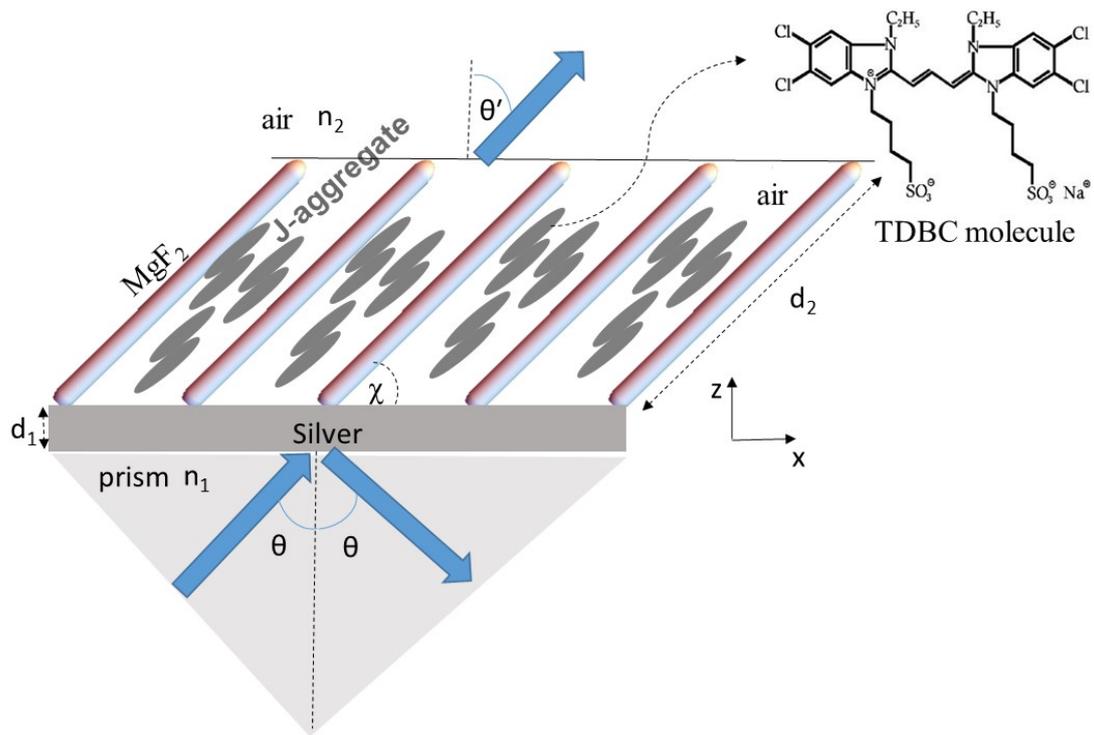

**Fig.1.** Schematic representation of Kretschmann configuration for excitation of surface plexciton at a metal/columnar thin film infiltrated with J-aggregate dyes interface. The chi (χ) is tilt angle of nanocolumns of columnar thin film.

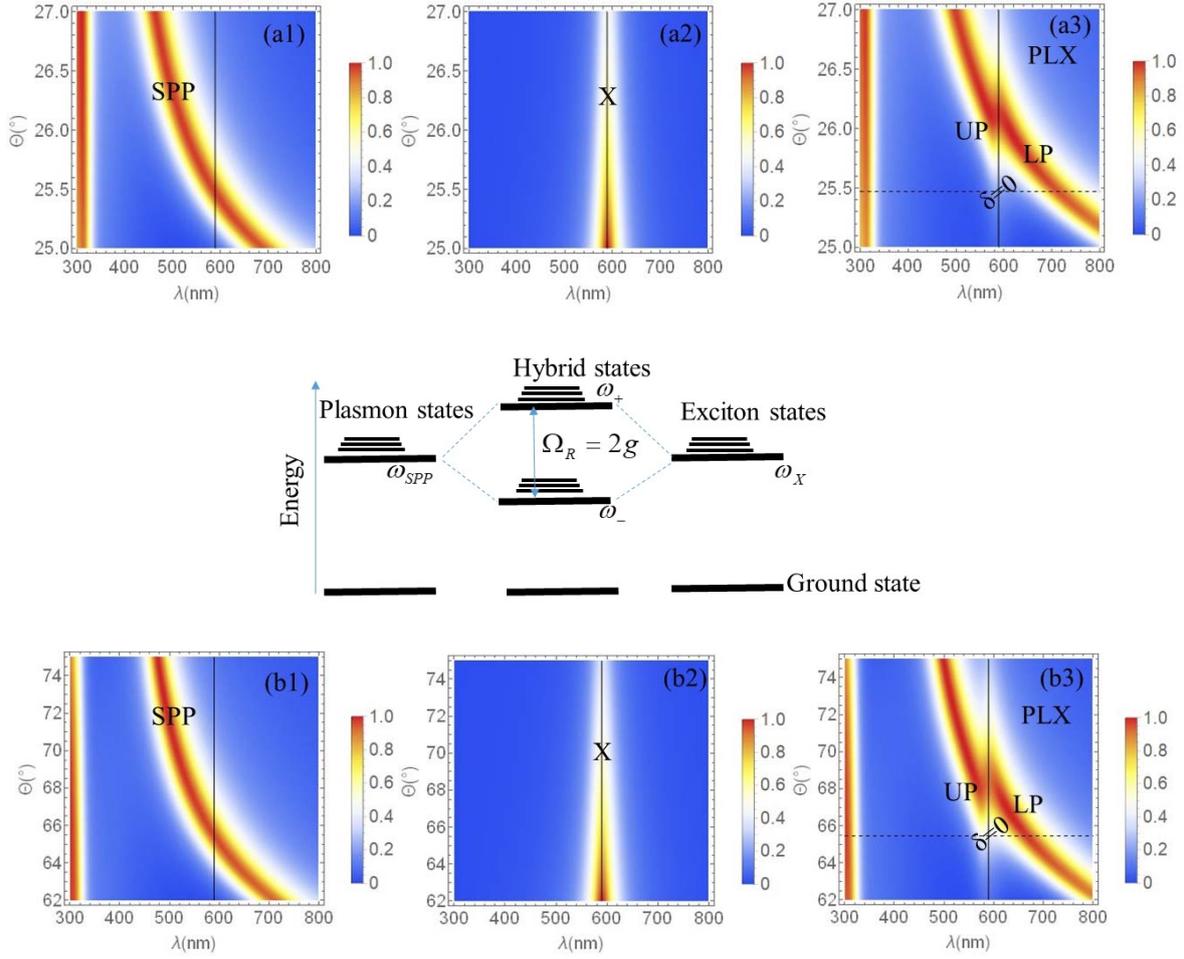

**Fig.2.** Absorption density plots as functions of wavelength and incident angle for excitation of surface plasmon polariton(SPP), surface exciton(X) and surface plexciton(PLX). Vertical lines in plots of SPP and X indicate to where detuning frequency is zero. In plexciton plot exist an anticrossing mode between LP and UP branches of polaritons at δ=0 so that horizontal line shows the location of incident angle at zero detuning frequency. The configuration in a1, a2 and a3 is described by the following parameters: $d_1$= 40 nm, $d_2$ = 800 nm, $n_1$=2.57, $n_2$ =1, $\chi$ =25 °, $f_j$ = 0.052, $f_v$ = 0.748, $\theta_\delta$ =25. 47°.These parameters in b1, b2, b3 are considered as $d_1$= 40 nm, $d_2$ = 800 nm, $n_1$=1.52, $n_2$ =1, $\chi$ =25 °, $f_j$ = 0.1, $f_v$ = 0.1, $\theta_\delta$ =65. 45°.

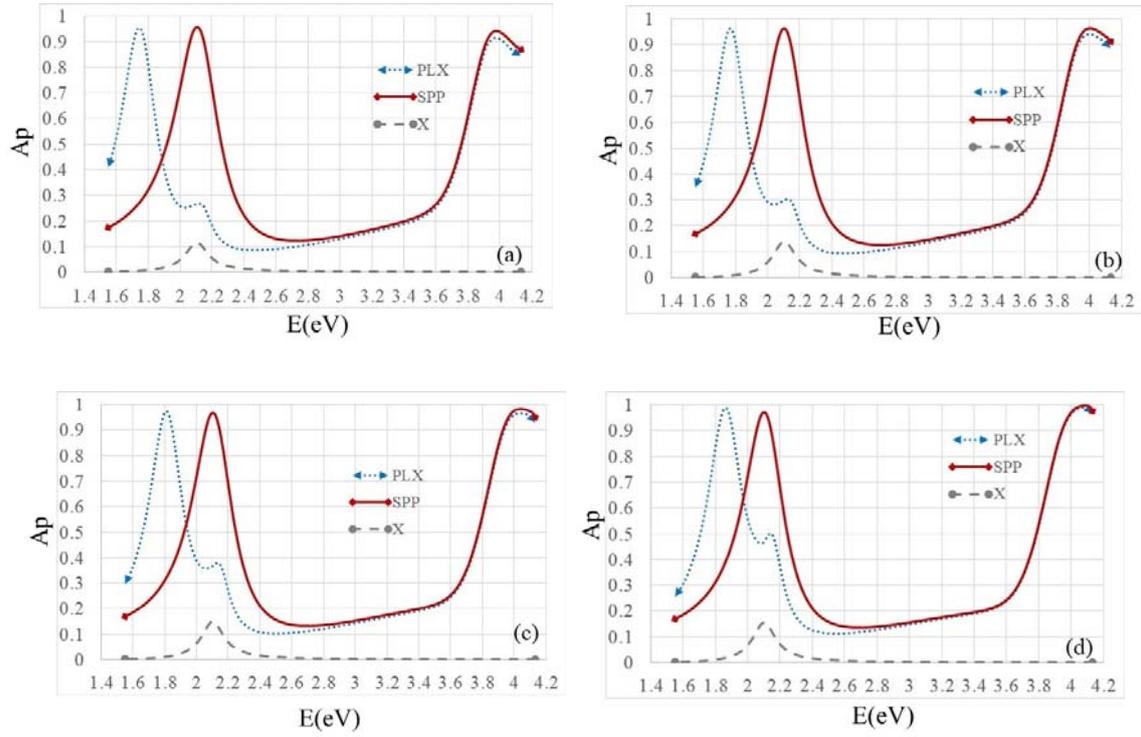

**Fig.3.** Optical absorption as a function of photon energy at different volume fractions of air and J-aggregate dyes with $d_1$= 40 nm, $d_2$ = 800 nm, $n_1$=2.57, $n_2$ =1, $\chi$ =25 ° and zero detuning frequency $\delta$=0  for excitation of surface exciton(X), surface plasmon polariton(SPP) and surface plexciton(PLX) (a) $f_j$ = 0.052, $f_v$ = 0.748, $\theta_\delta$=25.47° ($E_X$ = $E_{SPP}$ =2.105 eV), (b) $f_j$ = 0.068, $f_v$ = 0.523, $\theta_\delta$ =27.54°($E_X$ = $E_{SPP}$ =2.101 eV), (c) $f_j$ = 0.084, $f_v$ = 0.316, $\theta_\delta$ =29.91°($E_X$ = $E_{SPP}$ =2.101 eV) and (d) $f_j$ = 0.1, $f_v$ = 0.1 $\theta_\delta$=32.46°($E_X$ = $E_{SPP}$ =2.101eV).

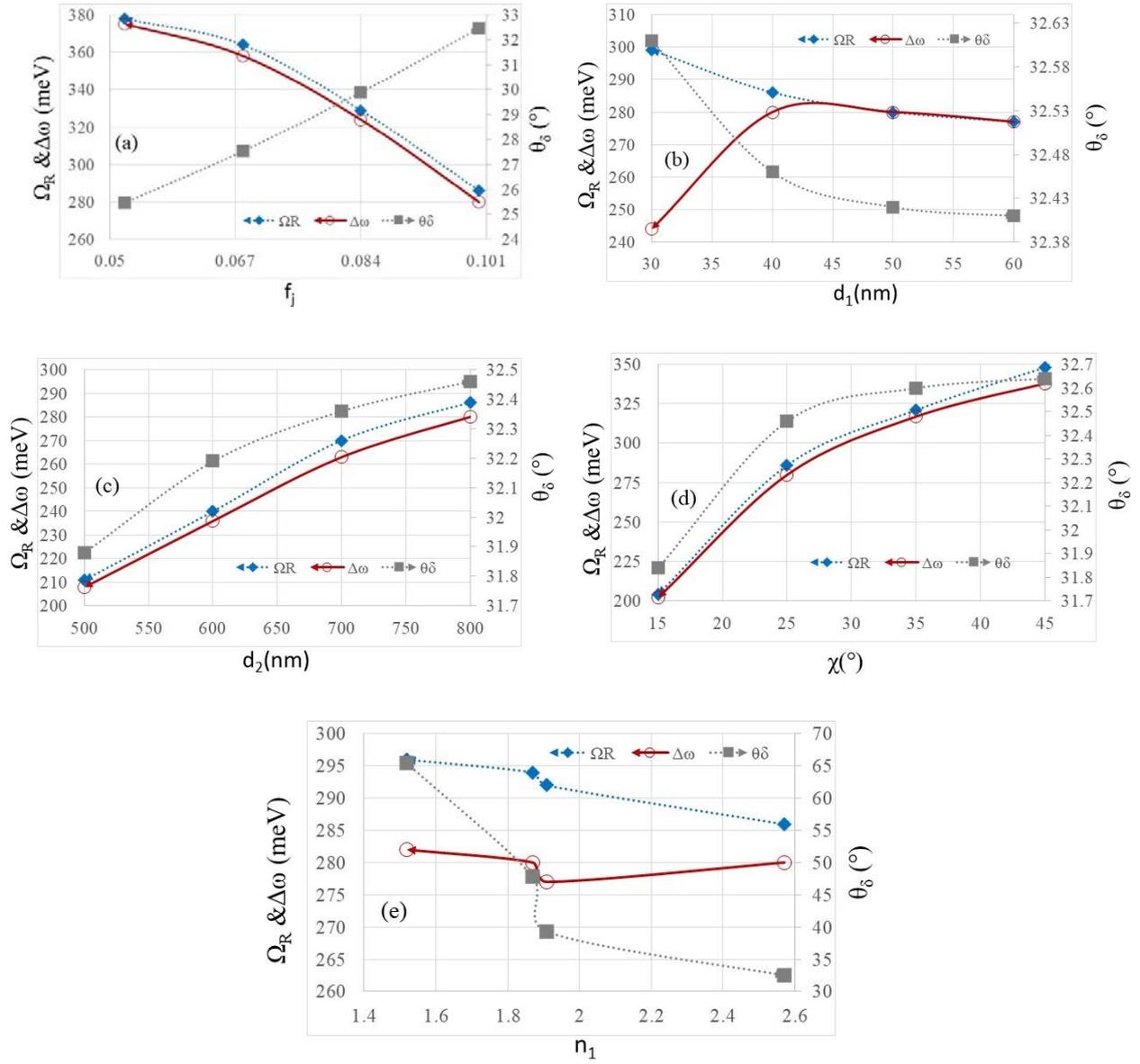

**Fig.4.** Rabi splitting energy $\Omega_R$, splitting energy $\Delta\omega$ and zero detuning angle $\theta_\delta$ (a) as a function of volume fractions J-aggregate dye, (b) as a function of thickness of metal, (c) as a function of length of exciton medium, (d) as a function of tilt angle of nanocolumns of exciton medium and (e) as a function of refractive index of prism. The other parameters are the same as in Fig.2

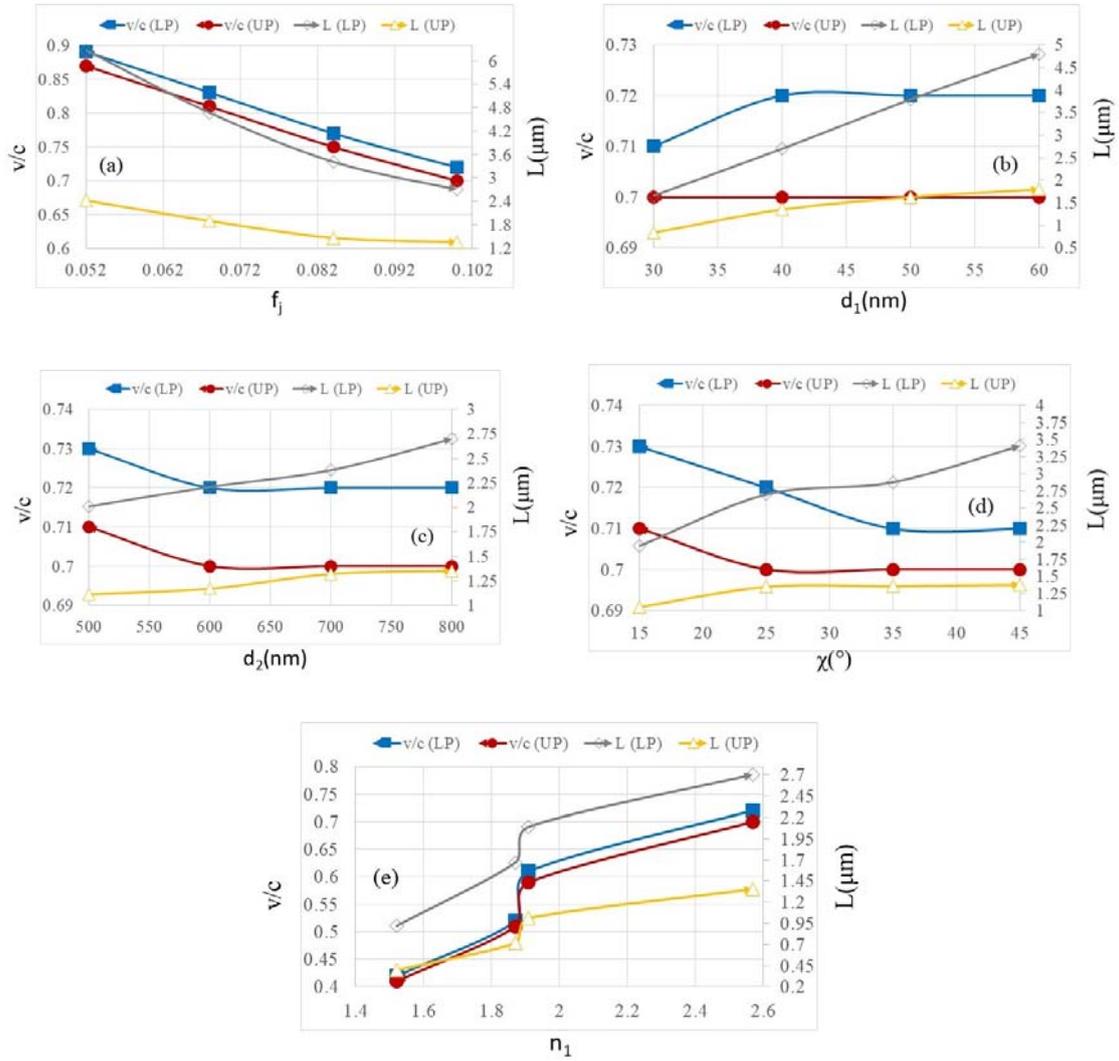

**Fig.5.** The relative phase velocity and propagation length of surface LP and UP plexciton wave (a) as a function of volume fractions J-aggregate dye, (b) as a function of thickness of metal, (c) as a function of length of exciton medium, (d) as a function of tilt angle of nanocolumns of exciton medium and (e) as a function of refractive index of prism. For the other parameters see Figs.S6 - S10 in supplementary material.

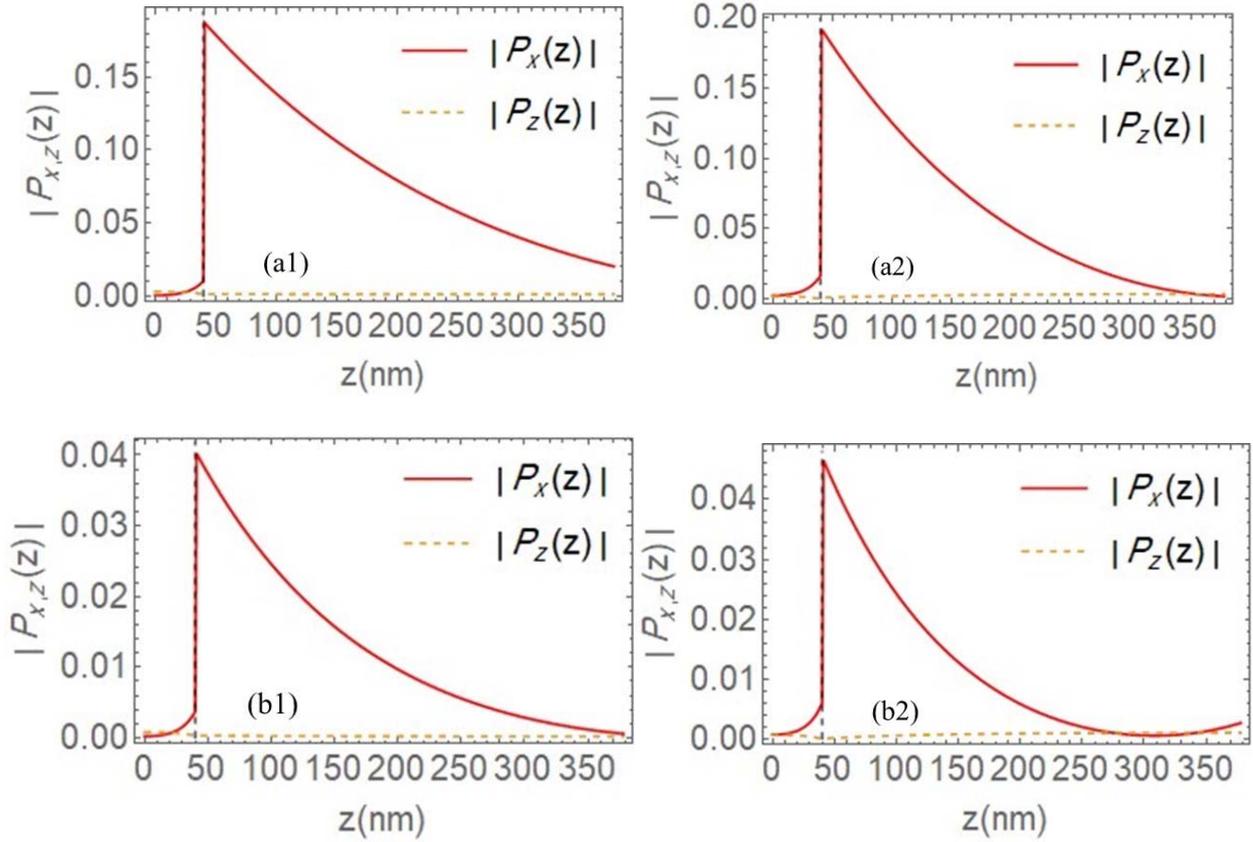

**Fig.6.** The plots show Cartesian components of time-averaged Poynting vector $|p_{x,z}(z)|$ along the z axis when a surface LP- and UP- plexciton waves have been excited at interface with zero detuning frequency with $d_1$= 40 nm, $d_2$ = 800 nm, $n_2$ =1, $\chi$ =25 ° .(a1)LP plexciton wave with fixed parameters : $n_1$=2.57, $f_j$ = 0.052, $f_v$ = 0.748, $E_{LP}$=1.744 eV,(a2) same as Fig.6 a1 except for UP plexciton wave and $E_{UP}$=2.119eV,(b1) LP plexciton wave with fixed parameters : $n_1$=1.52, $f_j$ = 0.1, $f_v$ = 0.1, $E_{LP}$ =1.859 eV and(b2) same as Fig.6 b1 except for UP plexciton wave and $E_{UP}$=2.141eV.

# Supplementary Material for "*Excitation* of surface plexciton wave at interface of a metal and a columnar thin film infiltrated with J-aggregate dyes"

F. Babaei [*], M. Rostami

*Department of Physics, University of Qom, Qom, Iran*

*)Email: *fbabaei@qom.ac.ir*

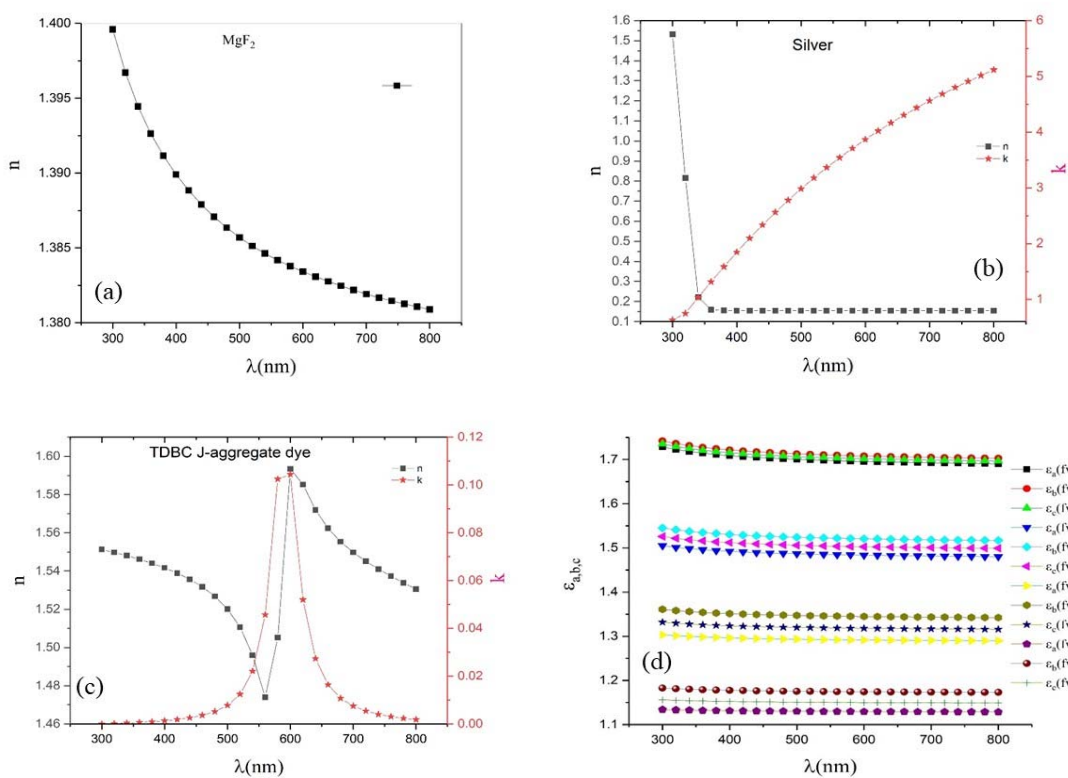

Fig.S1. Refractive indexes (n, k) of $MgF_2$ [1] (a), Ag [2] (b) and J-aggregate TDBC dyes [3] (c). (d) Dielectric permittivity scalars $\varepsilon_{a,b,c}$ ($\lambda$, $f_v$) of porous $MgF_2$ columnar thin film with different void volume fractions.

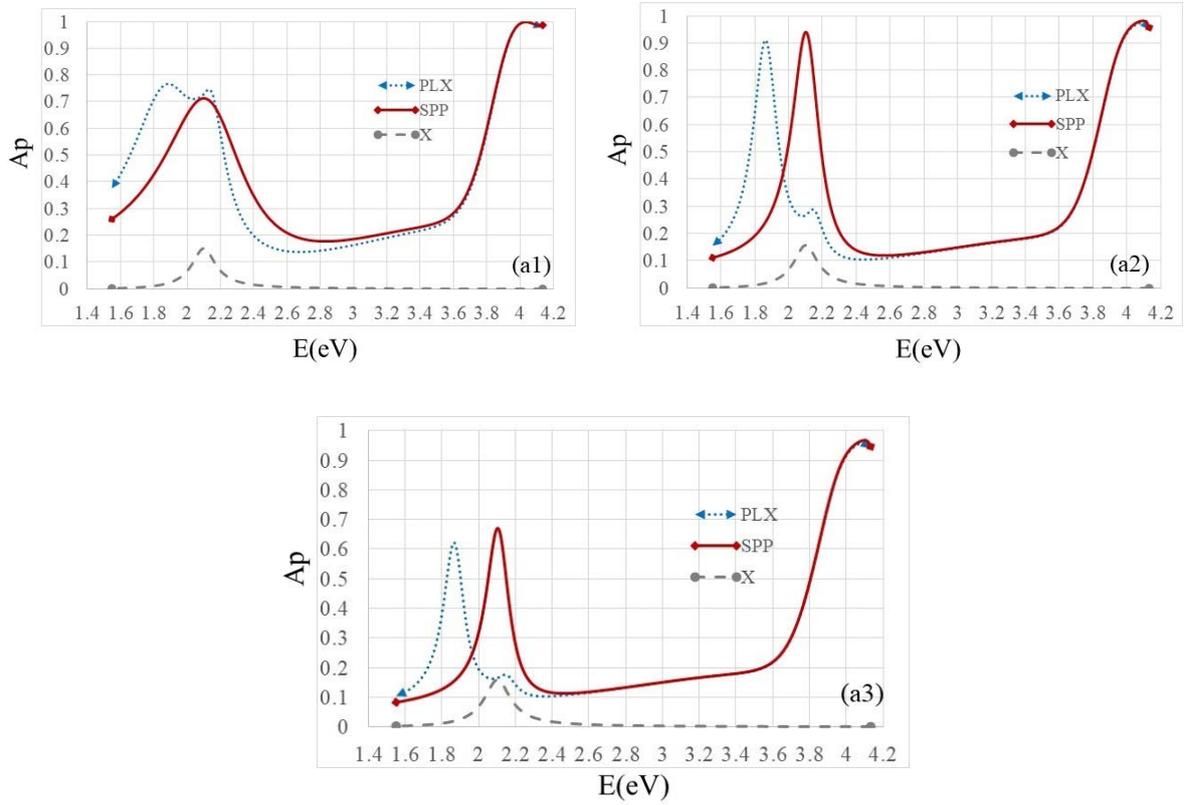

Fig.S2. Optical absorption as a function of photon energy for excitation of surface exciton(X), surface plasmon polariton(SPP) and surface plexciton(PLX) at different thicknesses of metal(Silver) with $f_v = 0.1$, $f_j = 0.1$, $d_2 = 800$ nm, $n_1 = 2.57$, $n_2 = 1$, $\chi = 25°$ and zero detuning frequency $\delta = 0$ (a1) $d_1 = 30$ nm, $\theta_\delta = 32.61°$ ($E_X = E_{SPP} = 2.101$ eV), (a2) $d_1 = 50$ nm, $\theta_\delta = 32.42°$ ($E_X = E_{SPP} = 2.101$ eV) and (a3) $d_1 = 60$ nm, $\theta_\delta = 32.41°$ ($E_X = E_{SPP} = 2.101$ eV).

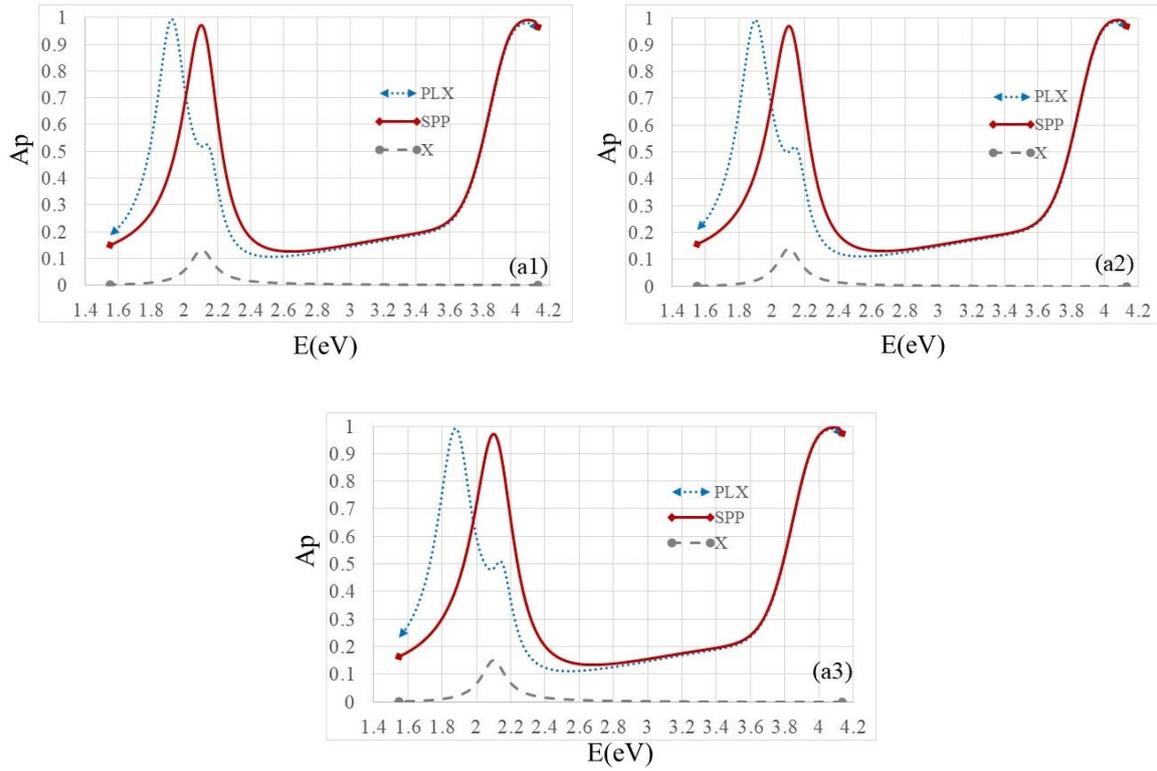

Fig.S3. Same as Fig.S2, except the Optical absorption was depicted at different lengths of exciton medium with $d_1$=40 nm (a1) $d_2$=500 nm, $\theta_\delta$=31.88°($E_X$ = $E_{SPP}$ =2.101eV),(a2) $d_2$=600 nm, $\theta_\delta$=32.19°($E_X$ = $E_{SPP}$ =2.101eV) and (a3) $d_2$=700 nm, $\theta_\delta$=32.36°($E_X$= $E_{SPP}$ =2.101eV).

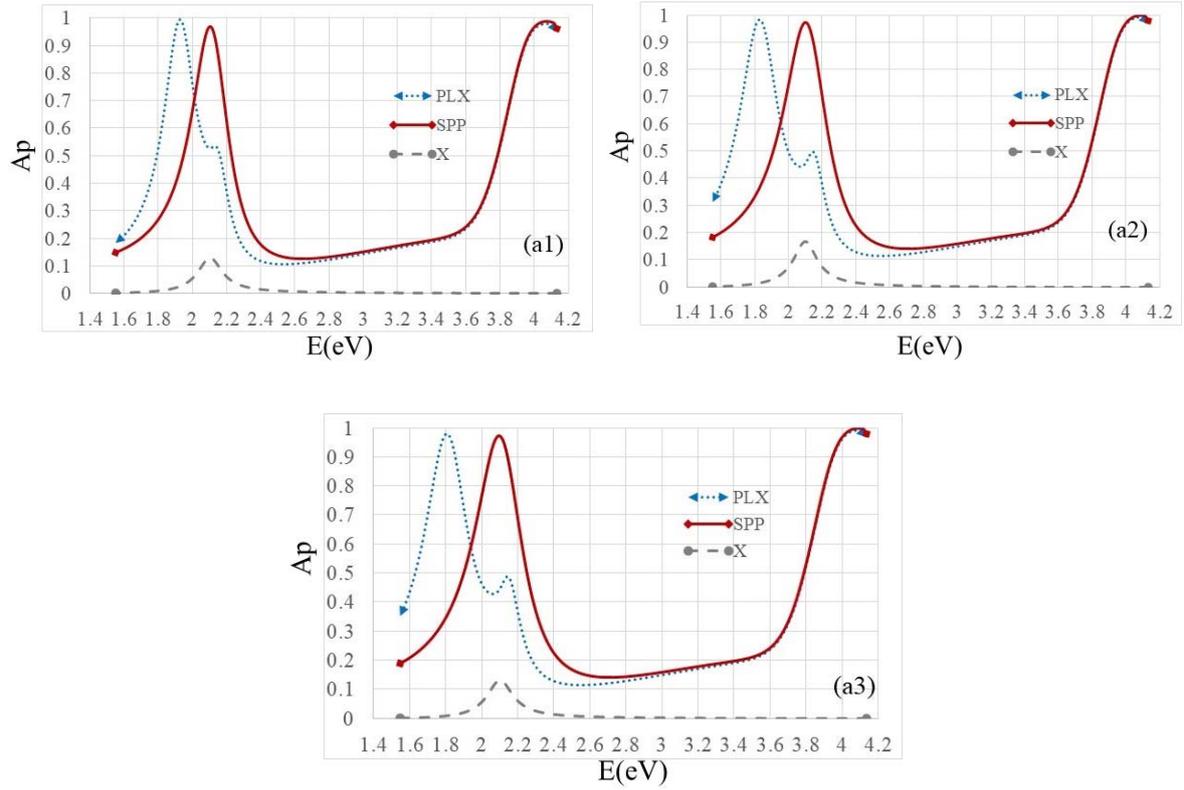

Fig.S4. Same as Fig.S2, except the Optical absorption was depicted at different tilt angles of nanocolumns of exciton medium with $d_1$=40 nm (a1) $\chi$=15°, $\theta_\delta$=31.84°($E_X = E_{SPP}$ =2.101eV),(a2) $\chi$=35°, $\theta_\delta$=32.6°($E_X = E_{SPP}$ =2.101eV) and (a3) $\chi$=45°, $\theta_\delta$=32.64°($E_X= E_{SPP}$ =2.098eV).

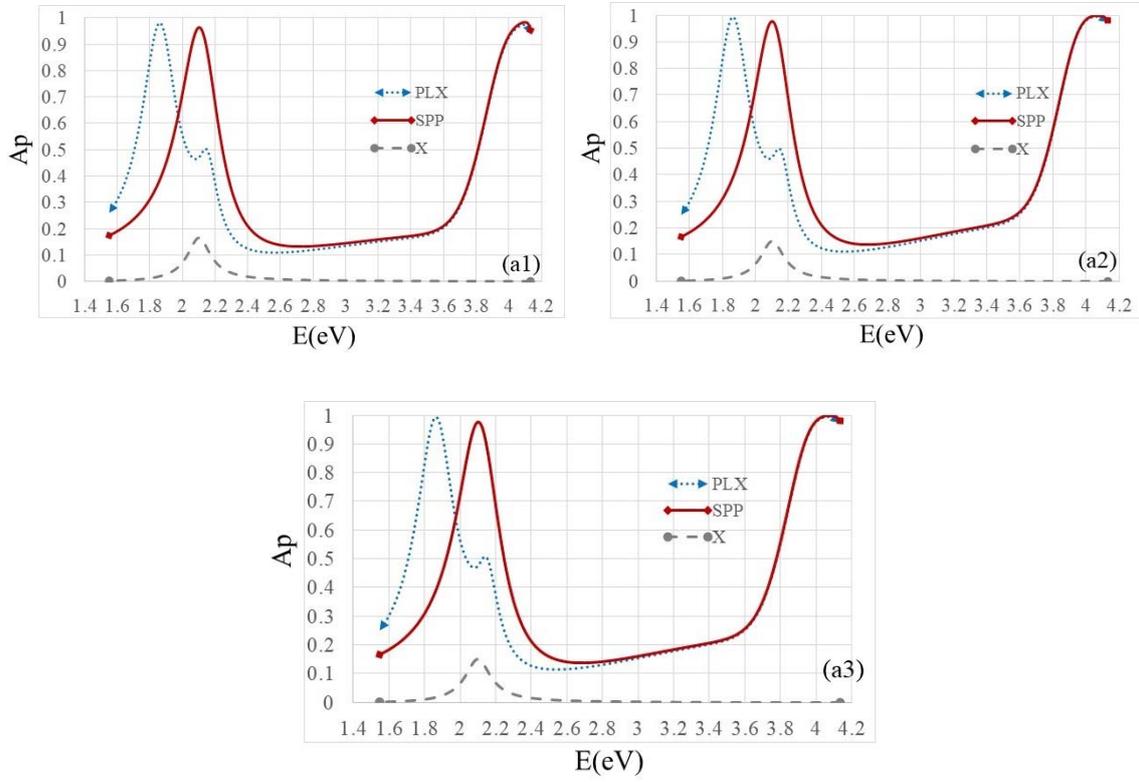

Fig.S5. Same as Fig.S2, except the Optical absorption was depicted at different prisms with $d_1$=40 nm (a1) $n_1$=1.52, $\theta_\delta$=65.45°($E_X$ = $E_{SPP}$ =2.101eV),(a2) $n_1$=1.87, $\theta_\delta$=47.82°($E_X$ = $E_{SPP}$ =2.101eV) and (a3) $n_1$=2.19, $\theta_\delta$=39.24°($E_X$= $E_{SPP}$ =2.098eV).

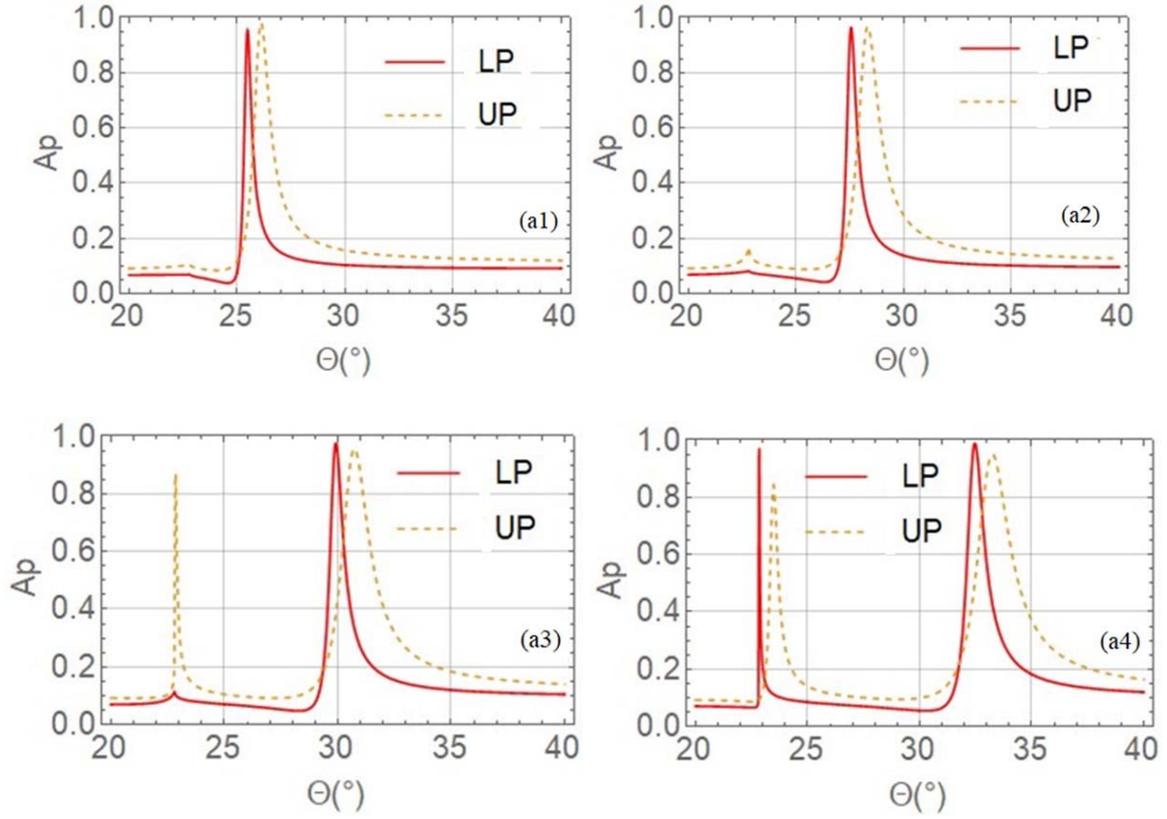

Fig.S6. Optical absorption as a function of incident angle at different volume fractions of air and J-aggregate dyes with $d_1$=40nm, $d_2$= 800 nm, $n_1$=2.57, $n_2$=1, $\chi$=25° was depicted at LP and UP peaks of plexciton plot (a1) $f_j$ = 0.052, $f_v$ = 0.748, $E_{LP}$ =1.744 eV, $E_{UP}$ =2.119 eV, (a2) $f_j$ = 0.068, $f_v$ = 0.523, $E_{LP}$ = 1.768 eV, $E_{UP}$ = 2.126 eV, (a3) $f_j$ = 0.084, $f_v$ = 0.316, $E_{LP}$ = 1.810 eV, $E_{UP}$ =2.134 eV and (a4) $f_j$ = 0.1, $f_v$ = 0.1, $E_{LP}$ =1.861 eV, $E_{UP}$ =2.141eV.

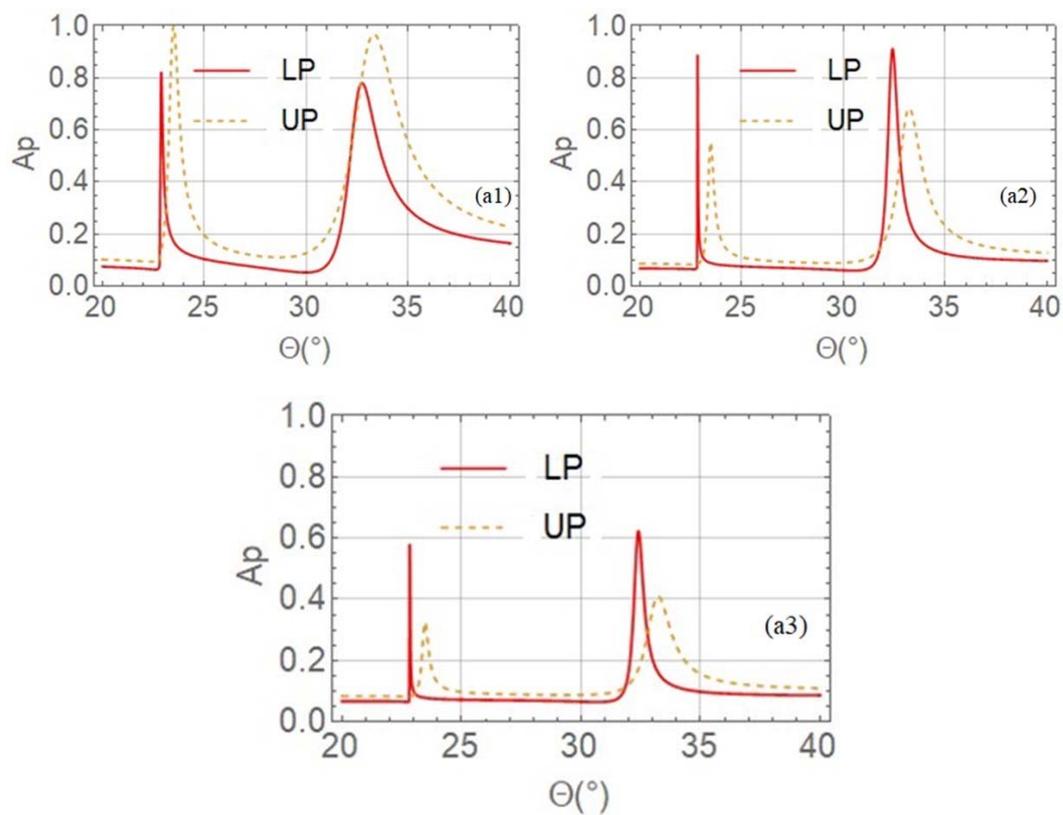

Fig.S7. Same as Fig.S6, except that the optical absorption was depicted at different thicknesses of metal(Silver) with $f_v$ = 0.1, $f_j$ =0.1 (a1) $d_1$=30 nm, $E_{LP}$=1.890 eV, $E_{UP}$=2.134 eV (a2) $d_1$=50 nm, $E_{LP}$=1.861 eV, $E_{UP}$=2.141 eV and (a3) $d_1$=60 nm, $E_{LP}$= 1.864eV, $E_{UP}$=2.141 eV.

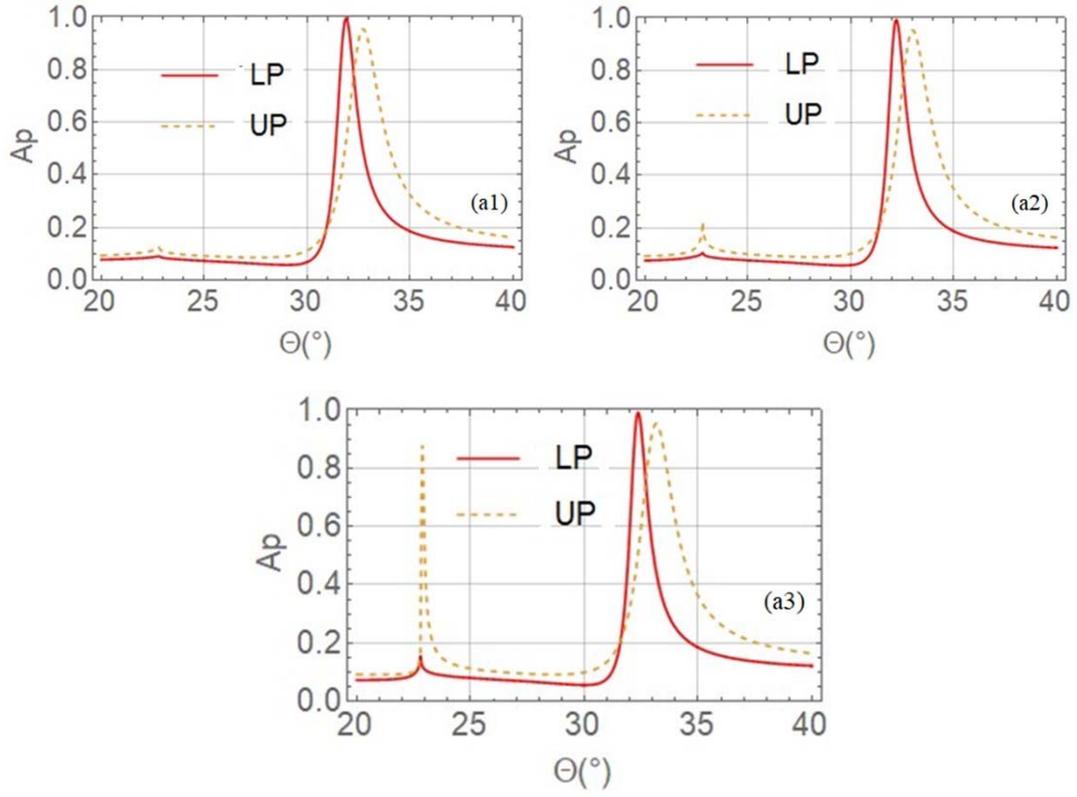

Fig.S8. Same as Fig.S6, except that the optical absorption was depicted at different lengths of exciton with $f_v = 0.1$, $f_j = 0.1$ (a1) $d_2$=500 nm, $E_{LP}$=1.922 eV, $E_{UP}$=2.130 eV (a2) $d_2$=600 nm, $E_{LP}$=1.901 eV, $E_{UP}$=2.137 eV and (a3) $d_2$=700 nm, $E_{LP}$= 1.878eV, $E_{UP}$=2.141 eV.

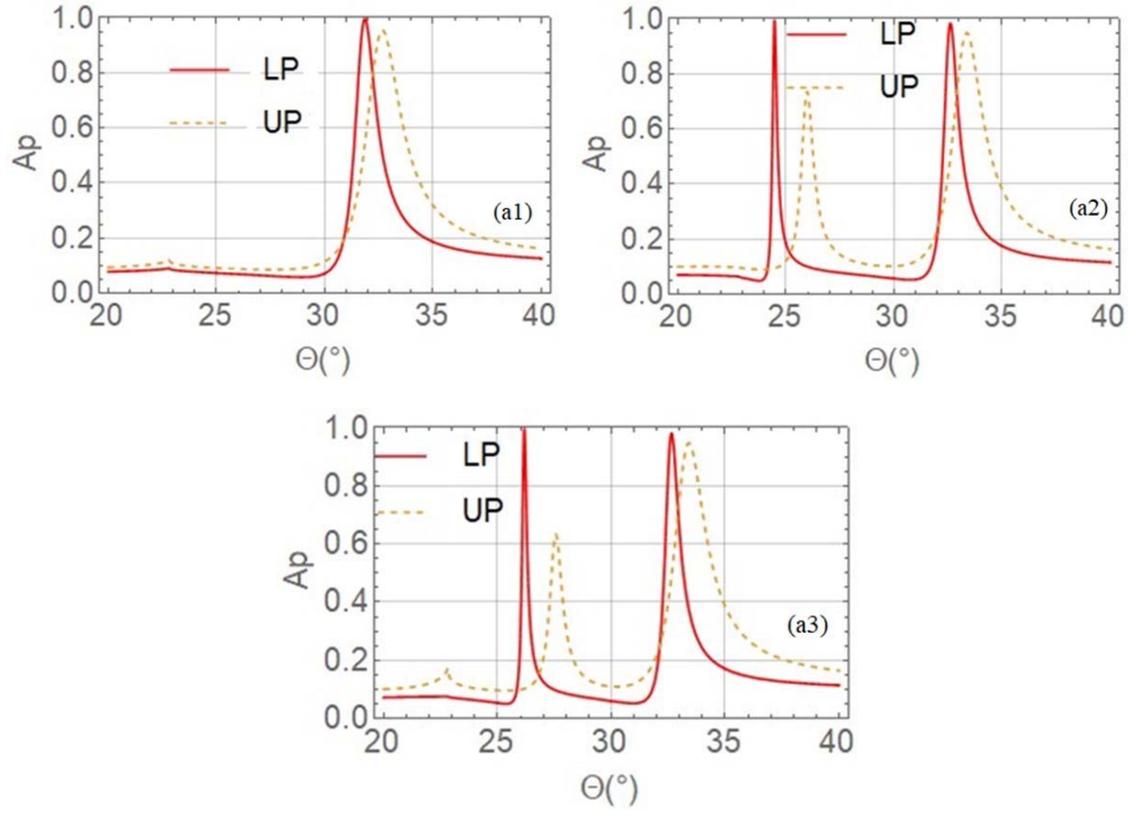

Fig.S9. Same as Fig.S6, except that the optical absorption was depicted at different tilt angles of nanocolumns of exciton medium with $f_v = 0.1$, $f_j = 0.1$ (a1) $\chi=15°$, $E_{LP}=1.928$ eV, $E_{UP}=2.130$ eV (a2) $\chi=35°$, $E_{LP}=1.828$ eV, $E_{UP}=2.145$ eV and (a3) $\chi=45°$, $E_{LP}=1.807$ eV, $E_{UP}= 2.145$eV.

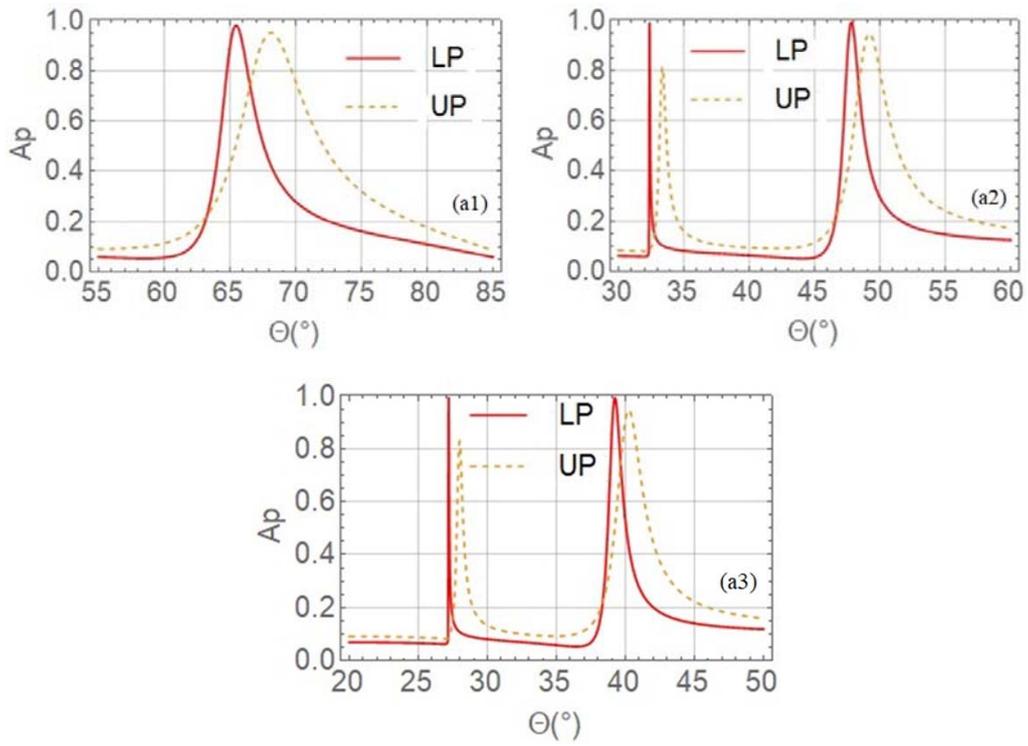

Fig.S10. Same as Fig.S6, except that the optical absorption was depicted at different prisms with $f_v = 0.1$, $f_j = 0.1$ (a1) $n_1=1.52$, $E_{LP}=1.859$ eV, $E_{UP}=2.141$ eV (a2) $n_1=1.87$, $E_{LP}=1.861$ eV, $E_{UP}=2.141$ eV and (a3) $n_1=2.19$, $E_{LP}=1.864$ eV, $E_{UP}=2.141$ eV.